\documentclass[aps,prb,twocolumn,groupedaddress,amsmath,amssymb,showpacs]{revtex4-2}

\usepackage{color}

\usepackage{graphicx}
\usepackage{hyperref}
\usepackage{pifont}
\usepackage{epstopdf}

\begin{document}
\title{Fe- and Co-based magnetic tunnel junctions with AlN and ZnO spacers}

\author{Gokaran Shukla$^1$}
\author{Stefano Sanvito$^2$}
\author{Geunsik Lee$^1$}
\affiliation{${}^1$Department of Chemistry, Center for Superfunctional Materials, Ulsan National Institute of Science and Technology, Ulsan 44919, Republic of Korea}
\affiliation{${}^2$School of Physics, AMBER and CRANN Institute, Trinity College, Dublin 2, Ireland}

\date{\today}


\begin{abstract}
AlN and ZnO, two wide band-gap semiconductors extensively used in the display industry, crystallise 
in the wurtzite structure, which can favour the formation of epitaxial interfaces to close-packed common 
ferromagnets. Here we explore these semiconductors as material for insulating barriers in magnetic 
tunnel junctions. In particular, the {\it ab initio} quantum transport code {\it Smeagol} is used to model 
the $X$[111]/$Y$[0001]/$X$[111] ($X=$~Co and Fe, $Y=$~AlN and ZnO) family of junctions.
Both semiconductors display a valance-band top with $p$-orbital character, while the conduction band 
bottom exhibits $s$-type symmetry. The smallest complex-band decay coefficient in the forbidden 
energy-gap along the [0001] direction is associated with the $\Delta_1$ symmetry, and connects across 
the band gap at the $\Gamma$ point in 2D Brillouin zones. This feature enables spin filtering and may 
result in a large tunnelling magnetoresistance. In general, we find that Co-based junctions present limited 
spin filtering and little magnetoresistance at low bias, since both spin sub-bands cross the Fermi level with 
$\Delta_1$ symmetry. This contrasts the situation of Fe, where only the minority $\Delta_1$ band is available. 
However, even in the case of Fe the magnitude of the magnetoresistance at low bias remains relatively small, 
mostly due to conduction away from the $\Gamma$ point and through complex bands with symmetry different 
than $\Delta_1$. The only exception is for the Fe/AlN/Fe junction, where we predict a magnetoresitance of 
around 1,000\% at low bias.
\end{abstract}

\maketitle

\section{Introduction}

The magnetoresistance effect represents the backbone of many spin-based devices~\cite{spintronics}, enabling the function of 
magnetic random-access memories~\cite{Bhatti2017}, sensors~\cite{Fujiwara2018}, spin-transfer-torque devices~\cite{Ralph2008}, 
microwave-generators~\cite{Kiselev2003}, and next-generation spin-based neuromorphic computing~\cite{Torrejon2017,Dutta2021}. 
The most prototypical device exploiting magnetoresistance is the magnetic tunnel junction (MTJ), where two ferromagnetic electrodes
are separated by an insulating barrier. This can operate as a binary unit, since typically its electrical resistance is minimal when the 
magnetization vectors of the two electrodes are parallel to each other, while it is maximised for an anti-parallel orientation. The magnitude
of the MTJ sensitivity is conventionally measured by the tunnelling magnetoresistance (TMR) ratio, defined as 
TMR~$=(R_\mathrm{AP}-R_\mathrm{P})/R_\mathrm{P}$, where $R_\mathrm{P}$ and $R_\mathrm{AP}$ are the resistances of the MTJ 
in parallel and antiparallel configuration, respectively. 

Although the magnetic data storage industry was revolutionarized first by the giant magnetoresistance 
effect in metallic magnetic multilayers~\cite{Fert1988,Grunberg1989}, TMR-based MTJs today represent 
the state-of-the-art technology, owing to their large TMR ratios, reaching up to 200~\% at room 
temperature~\cite{Parkin2004,Yuasa2004}. Early MTJ devices were based on amorphous 
tunnelling barriers, mostly Al$_2$O$_3$ \cite{Miyazaki,Moodera}, for which the magnitude
of the TMR is determined by the spin polarisation of the density of states (DOS)~\cite{Mazin}, 
$P=\frac{n_{\uparrow}-n_{\downarrow}}{n_{_\uparrow}+n_{_\downarrow}}$, where  
$n_{\uparrow}$ ($n_\downarrow$) is the spin-up (down) DOS at Fermi energy, $E_\mathrm{F}$.
For these structures the TMR ratio can be estimated by Julliere's relation, 
TMR~$=\frac{2P_1 P_2}{1-P_1 P_2}$, where $P_1$ and $P_2$ are the DOS spin polarisations of 
the ferromagnetic electrodes~\cite{Julliere}. Since in transition metals $P$ hardly exceeds 50\%, 
the expected TMR ratios for amorphous barriers remain limited. A different situation, however,
is encountered for epitaxial MTJs, where the transverse wave-vector $ \bf k_{||}$ is conserved 
during tunnelling thus remaining a good quantum number. The tunnelling probability is then
determined by the symmetry of the wave-function. As this can be different for the two different 
sub-bands of a magnetic metal, spin filtering is expected and hence arbitrary large TMR 
ratios~\cite{Butler,Mathon}. Such spin-filtering effect has been confirmed 
experimentally~\cite{Parkin2004,Yuasa2004} and it is at the foundation of modern high-performance
TMR-based devices. Inspired by these initial works a multitude of materials compositions offering spin
filtering have been proposed~\cite{Velev2009,Nuala2012,Jutong2012,Faleev2015,Heusler,shukla}.

Interestingly, although the symmetry filtering argument is applicable to many all-epitaxial junctions,
only a particular stack has shown its potential in the real world, namely the Fe/MgO/Fe MTJ. There
are several arguments in favour of Fe/MgO: 1) an epitaxial growth with strong suppression of the 
interface defects, which arise due to the lattice mismatch between the metal and the insulator; 
2) a well-consolidated growth recipe, which can scale up to large surface areas; 3) a large 
in-plane/perpendicular magnetocrystalline anisotropy for FeCoB magnetic electrodes; 4) the 
robust and wide band-gap of MgO, which ensures ideal tunnelling. Nonetheless, the FeCoB/MgO 
system also presents some disadvantages at the fabrication and operation level. In particular, the 
growth a typical Fe/MgO-based MTJ requires several layers of lithographic process for different 
materials with various optimal thickness to pin the reference layer magnetic moments in a certain 
direction. 

In fact, most ferromagnetic materials crystallize with a sixfold rotation symmetry ($C_6$), whereas 
insulating barrier materials, such as  MgO, are only fourfold ($C_4$). In general, it is difficult to 
grow epitaxially $C_4$ MgO on $C_6$ substrates with a minimum interface vacancy content. 
For this reason, and for opening up the avenue to new classes of devices, it becomes interesting 
to explore whether high-performing MTJs with sixfold rotation symmetry can be made. This is the 
task set out for our work, which investigates a family of MTJs constructed with the wide-gap wurtzite 
insulators, AlN and ZnO. These are widely used as light-emitting-diode materials in the 
microelectronics industry. Should they work at polarizing the current, one may also imagine 
the possibility of realizing spin-polarized current-based displays with circular-polarized light for 
high-viewing angle~\cite{Holub}.
      
The paper is organized as follows. In the next section we present first our computational method 
and the details of the present work. Then, we discuss our calculated real band structures of the 
ferromagnetic electrodes and the complex bands of the insulating barriers, before moving to 
an analysis of the transmission coefficients and the associated TMR. Finally we conclude.

\section{Computational details}
The electronic structure of the various materials forming our MTJs is calculated with the density 
functional theory (DFT) formalism using the {\sc Siesta} code~\cite{soler-siesta} . {\sc Siesta} employs 
norm-conserving pseudo-potentials and a numerical atom-centered local-orbital basis set. The 
many-body interacting problem is solved through an auxiliary effective single-body non-interacting 
Kohn-Sham potential, where the exchange-correlation functional is treated at the level of local 
density approximation (LDA) with the Ceperly-Alder parameterization~\cite{Ceperly}. Quantum 
transport is computed with the non-equilibrium Green's function method, implemented within the 
Kohn-Sham DFT Hamiltonian (the so-called NEGF+DFT scheme) in the {\sc Smeagol} 
code~\cite{sanvito-smeagol1,sanvito-smeagol2,sanvito-smeagol3}. {\sc Smeagol} uses {\sc Siesta} 
as DFT engine.

The complex band-structures~\cite{Bosoni2022} of AlN and ZnO are calculated by taking [0001] as 
the transport direction ($z$-axis), and we restrict ourselves to the special lines with ${\bf k}_{||}=0$ 
($\bf k_{||}$ is the wave-vector in the plane transverse to the transport direction). This choice is
justified by the evaluation of the minimum complex decay coefficient along $z$ over the entire 
transverse Brillouin zone. In all cases we set the 
real mesh cutoff to 700~Ryd and take a 8$\times$8$\times$8 $k$-point mesh for the Monkhorst-Pack 
sampling. The Bloch orbitals are expanded with a basis set of double-$\zeta$ quality for the $s$, $p$ 
and $d$ shells of Co and Fe, while a double-$\zeta$ plus polarization one is employed for the $s$ 
and $p$  orbitals of Al, N, Zn and O.

We then design four different MTJs, namely Fe/AlN/Fe, Co/AlN/Co, Fe/ZnO/Fe and Co/ZnO/Co.
The experimental in-plane lattice constants of bulk AlN and ZnO are 3.09~\AA\ and 3.25~\AA, 
respectively. In order to obtain, commensurate junctions we adjust the in-plane lattice constants to 
3.34~\AA\  for both the insulators and take 2.73~\AA\ for \textit{bcc}-Fe and \textit{bcc}-Co. The 
interface is then formed by matching a 2$\times$2 (0001) surface of the semiconductors with a
3$\times$3 one for the metals. Thus, the hexagonal (0001) plane (lattice constant 6.69~\AA) 
of the semiconductor is epitaxial to the (111) one of Fe and Co after a 30$^{o}$ rotation about 
the [111] direction. This matching requires 8~\% and 6~\% tensile strain on AlN and ZnO, respectively.
At the same time Fe is under a compressive strain of about 3~\%. When we compare the electronic 
band-structures of the various materials at such lattice parameters we notice little qualitative 
variation due to the strain. This allows us to obtain quantitative results for junctions having a 
computationally-manageable cells. We  then relax all the atomic coordinates using conjugate 
gradient until the forces are smaller than ~0.01eV/\AA. After relaxation, the structure of both 
AlN and ZnO transforms from bulk wurtzite into a graphite like. 

In general, both ZnO- and AlN-based MTJs turn out to be symmetric about the plane located 
in the middle of layer. We select 12.5~\AA\ and 24.5~\AA\ thick barriers for AlN and ZnO, 
respectively.
  
Next we use {\sc Smeagol} to perform electron transport calculations. At a given bias voltage, $V$, 
{\sc Smeagol} calculates the electrical current, $I$, for both spins $\sigma$ ($\sigma=\uparrow, \downarrow$) 
using the Landauer-B\"uttiker coherent transport formalism as,
\begin{equation}\label{current}
I^{\sigma}(V)  = \frac{e}{h}\int dE\:  T^{\sigma}(E;V)[f_\mathrm{L}(E,\mu)-f_\mathrm{R}(E,\mu)]\:,
\end{equation}
where $e$ is the electron charge, $h$ the Plank constant, $T^{\sigma}(E;V)$ the energy- and bias-dependent transmission 
coefficient and $f_\mathrm{L}$ ($f_\mathrm{R}$) the Fermi function associated to the left-hand (right-hand) side electrode. 
This is evaluated at $E-\mu_\mathrm{L}$ ($E-\mu_\mathrm{R}$), where $\mu_\mathrm{L/R}=E_\mathrm{F}\pm\frac{eV}{2}$ 
is the chemical potential for the left/right electrode. Since the junction is translationally invariant over the $x-y$ plane (periodic 
boundary conditions), the transmission coefficient can be written over the 2D Brillouin zones as,
\begin{equation}\label{tkd}
T^{\sigma}(E;V) = \frac{1}{\Omega_\mathrm{BZ}}\int_\mathrm{BZ} d\mathbf{k}_\parallel\: T^{\sigma}_{\mathbf{k}_\parallel}(E;V)\:,
\end{equation}
where $\Omega_\mathrm{BZ}$ is the volume of the two dimensional first Brillouin zone. The $\mathbf{k}_\parallel$-dependent
transmission coefficient is then obtained as,
\begin{equation}
T^{\sigma}_{\mathbf{k}_\parallel}(E;V)=Tr[\Gamma_\mathrm{L}^{\sigma}(E;V) G^{\dagger \sigma}_\mathrm{C}(E;V)\Gamma_\mathrm{R}^{\sigma}(E;V)
G^{\sigma}_\mathrm{C}(E;V)]
\end{equation}
with the retarded Green's function of the scattering region given by 
$G^\sigma_\mathrm{C}(E;V)= \lim\limits_{\eta\rightarrow0}[E+i\eta -H_\mathrm{C} -\Sigma^\sigma_\mathrm{L}(E;V)-\Sigma^\sigma_\mathrm{R}(E;V)]^{-1}$, 
where $H_\mathrm{C}$ is Hamiltonian of device scattering region and the coupling matrices $\Gamma_\mathrm{L/R}^{\sigma}$ are 
related to the lead self-energy matrices by $\Gamma_\mathrm{L/R}^\sigma= \Sigma_\mathrm{L/R}^\sigma - \Sigma_\mathrm{L/R}^{\dagger\sigma}$.  
The transport calculation is then performed in the zero-bias limit with the electrons distribution converged on a 8$\times$8$\times$1 $k$-point 
grid, while the transmission coefficient integration is performed over 50$\times$50$\times$1 k-mesh. We have also carried out additional tests 
for a 100$\times$100$\times$1 mesh without noting any notable change in $T^{\sigma}(E;V)$ or in TMR. 

\section{Results and Discussion}

\subsection{AlN  and ZnO as tunnelling barriers}

AlN and ZnO are two wide band-gap semiconductors mostly used in the opto-electronics industry for light emitting diodes.
For AlN the LDA calculates a band-gap of 4.15~eV direct at the $\Gamma$ point, a value that is $\sim$2~eV smaller 
than the experimental one (6.1 eV). An even more severe underestimation is found for ZnO, with an LDA gap of 0.61~eV 
(direct at $\Gamma$) against the experimental measure of 3.25~eV. In order to overcome the well-known band gap underestimation 
problem of the LDA, we employ the atomic self-interaction correction (ASIC) scheme~\cite{SIC1,SIC2}. This returns us a band-gap
of $\sim$5.6 eV for AlN and $\sim$3.2 eV for ZnO, which are close to the aforementioned experimental values. The ASIC is then
used for the transport calculations.
  
We begin our investigation by computing the complex band-structure~\cite{Bosoni2022} of the two insulators used as tunnel barrier.
Recalling that $z$ is the direction of the electron transport, the conventional band equation, $E=\epsilon(\mathbf{k}_\parallel, k_z)$,
can be extended to energies, $E$, within the forbidden band-gap, by continuing $k_z$ over the complex axis, namely by taking 
$k_z=i\kappa$. Here, $\epsilon$ is a function of the wave-vector $(\mathbf{k}_\parallel, k_z)$ and so that $\kappa$ describes the
exponential decay of the wave-function for a given energy $E$ in the band-gap and a particular transverse wave-vector, 
$\mathbf{k}_\parallel$. The transmission coefficient across an insulating barrier of thickness $d$ will then be,
$T(E,\mathbf{k}_\parallel)\sim T_0(E,\mathbf{k}_\parallel)\mathrm{e}^{-2\kappa(E,\mathbf{k}_\parallel)d}$,
where $T_0(E,\mathbf{k}_\parallel)$ in general depends on the nature of the interface between the metal and the
insulator. The decay constant $\kappa(E,\mathbf{k}_\parallel)$ varies with the magnitude of transverse wave-vector and the orbital 
symmetry~\cite{Butler2} as $k=\sqrt{(\frac{2m}{\hbar^2}) (V-E)+k_{||}^2 -\frac{<\phi|(\frac{\partial}{dx^2}+ \frac{\partial}{dy^2}|\phi>)}{<\phi|\phi>}}$, 
where the last term (Laplacian) describes the transverse oscillation of tunnelling wave-functions.
One can then plot $\kappa(E_\mathrm{F},\mathbf{k}_\parallel)$ in 2D Brillouin zone spanned by the transverse wave-vector $\mathbf{k}_\parallel$ 
and establish which portions of the Brillouin zone contribute the most to the tunnelling current. The higher value of 
$\kappa(E_\mathrm{F},\mathbf{k}_\parallel)$ corresponds to smaller transmission probability amplitude.   

\begin{figure}[htp]
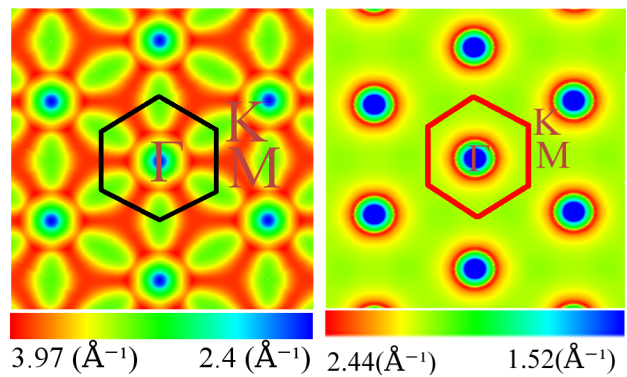

\centering
\includegraphics[width =4.1cm]{Fig1a.pdf}
\includegraphics[width =4.0cm]{Fig1b.pdf}
\caption{Heat colour plots of the wave-function decay coefficient, $\kappa(E_\mathrm{F},\mathbf{k}_\parallel)$, as a function of the 
transverse wave-vector, $\mathbf{k}_\parallel$, for AlN (left-hand side panel) and ZnO (right-hand side panel). Calculations are carried 
out for $E_\mathrm{F}$ placed in the middle of the band gap. The black and red boxes mark the 2D Brillouin zones and the colour code 
is blue to green to red as $\kappa$ gets larger. In both cases the decay coefficient is plotted in linear scale with the following limit: 
AlN $\kappa_\mathrm{min}=2.4~$\AA$^{-1}$, $\kappa_\mathrm{max}=3.97~$\AA$^{-1}$;
ZnO $\kappa_\mathrm{min}=1.52~$\AA$^{-1}$, $\kappa_\mathrm{max}=2.44~$\AA$^{-1}$. 
}  
\label{Fig1}
\end{figure}

This exercise is presented in Fig.~\ref{Fig1}, where we show the $\kappa(E_\mathrm{F},\mathbf{k}_\parallel)$ contour maps 
of AlN and ZnO over the first Brillouin zone of the 2D transverse hexagonal lattice.
The figure clearly shows that both AlN and ZnO exhibit the smallest wave-vector decay coefficient at the $\Gamma$ point. This is
the situation corresponding to a tunnelling electron approaching the barrier along the transport direction normal to the surface, namely 
when the  effective distance travelled by the electrons across the barrier region is minimal. Symmetry analysis further suggests that
the Bloch states available around the $\Gamma$ point have $\Delta_1$ symmetry. Comparing the two compounds, we found that AlN 
presents relatively large decay coefficients over the entire Brillouin zone, except for regions around  $\Gamma$, K and M, with $\Gamma$ 
being the most transmissive point in the Brillouin zone. In contrast for ZnO large transmission is found only at $\Gamma$, with little 
contribution from the rest of the $\mathbf{k}_\parallel$ plane. The decay constants are generally rather small, owing the large band gaps 
of these two compounds. Note that the distribution of the decay coefficients over the Brillouin zone is not expected to change with the 
choice of the DFT functional, which just modifies the band gap, but not the symmetry of the Kohn-Sham states.

Having established that most of the transmission is likely to take place at $\Gamma$, next we analyse in more detail the complex
band-structure along the transport direction for $\mathbf{k}_\parallel=0$. In epitaxial junctions, where $\mathbf{k}_\parallel$ is 
conserved, the largest contribution to the transmission coefficient in Eq.~(\ref{tkd}) originates from a region of the Brillouin
zone around $\mathbf{k}_\parallel=0$. Furthermore, one has to assign the symmetry of the tunnelling wave-function, since this
is also conserved in the phase-coherent tunnelling process. Such assignment is performed by projecting the wavefunction onto 
the transverse plane and by characterizing it according to its orbital composition. More specifically, a $\Delta_1$ symmetry is assigned 
to Bloch states  having zero angular momentum about the transport direction, the $z$ axis. This means that the $\Delta_1$ symmetry 
is associated to $s$, $p_z $ and  $d_{3z^{2}-r^{2}} $ orbitals. In contrast, the $p_x$, $p_y$, $d_{xz}$ and $d_{yz}$ orbitals are assigned 
to $\Delta_5$ symmetry, while $\Delta_2$ corresponds to the $d_{x^{2}-y^{2}}$ orbital. Finally, the $\Delta_{2^\prime}$ symmetry 
is characteristic of the $d_{xy}$ orbital.
\begin{figure}[htp]
\centering
\includegraphics[width =7.cm]{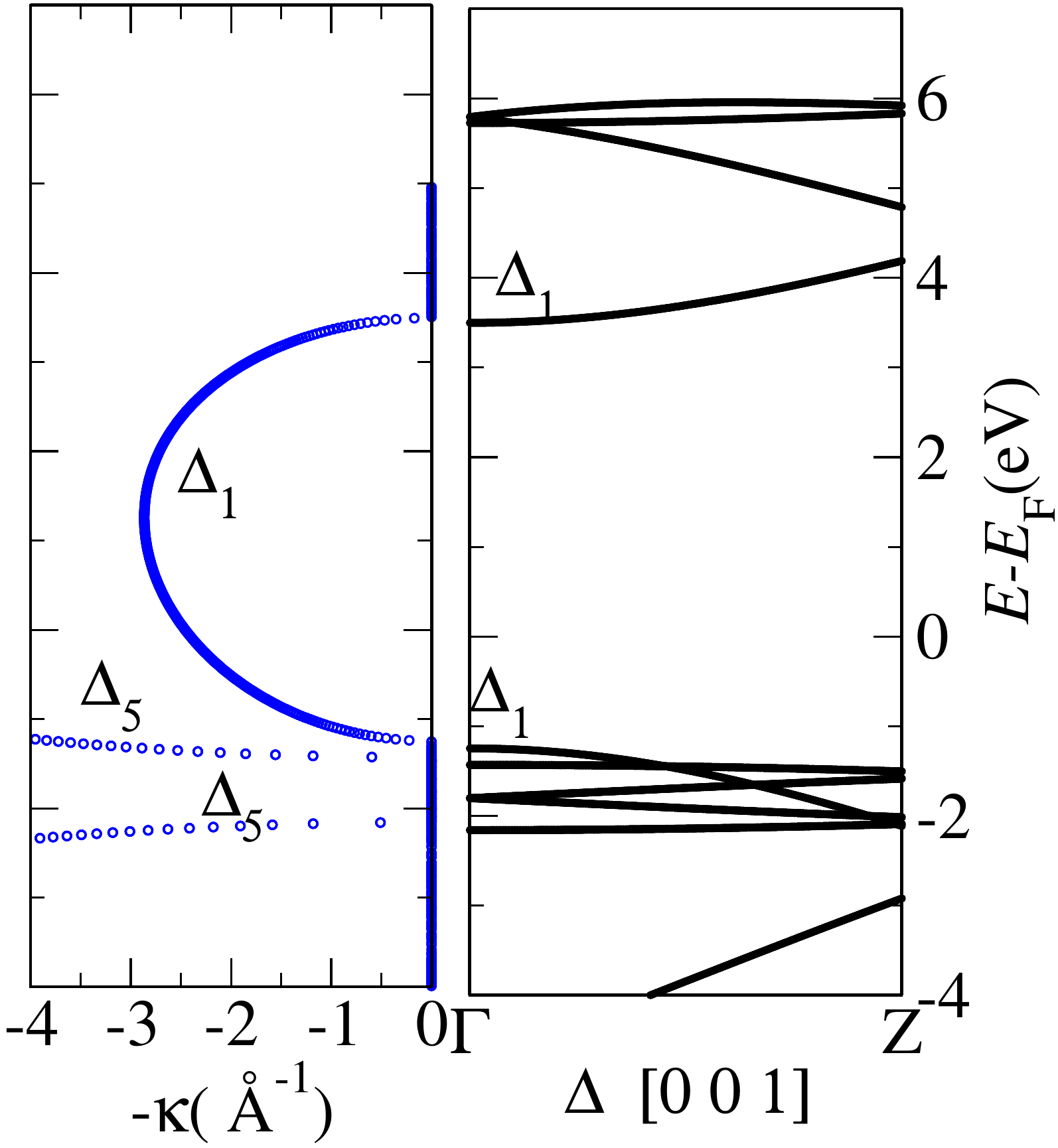}
\caption{Real (right-hand side panel) and complex (left-hand side panel) band structure of AlN calculated at the 
$\Gamma$ point in the 2D transverse Brillouin zone. The symmetry labels, $\Delta_n$, where n $\in$ [1,5], have 
been described in the text and the energy is measured from the Fermi energy $E_\mathrm{F}$.} 
\label{Fig3}
\end{figure}

\begin{figure}[htp]
\centering
\includegraphics[width =7.cm]{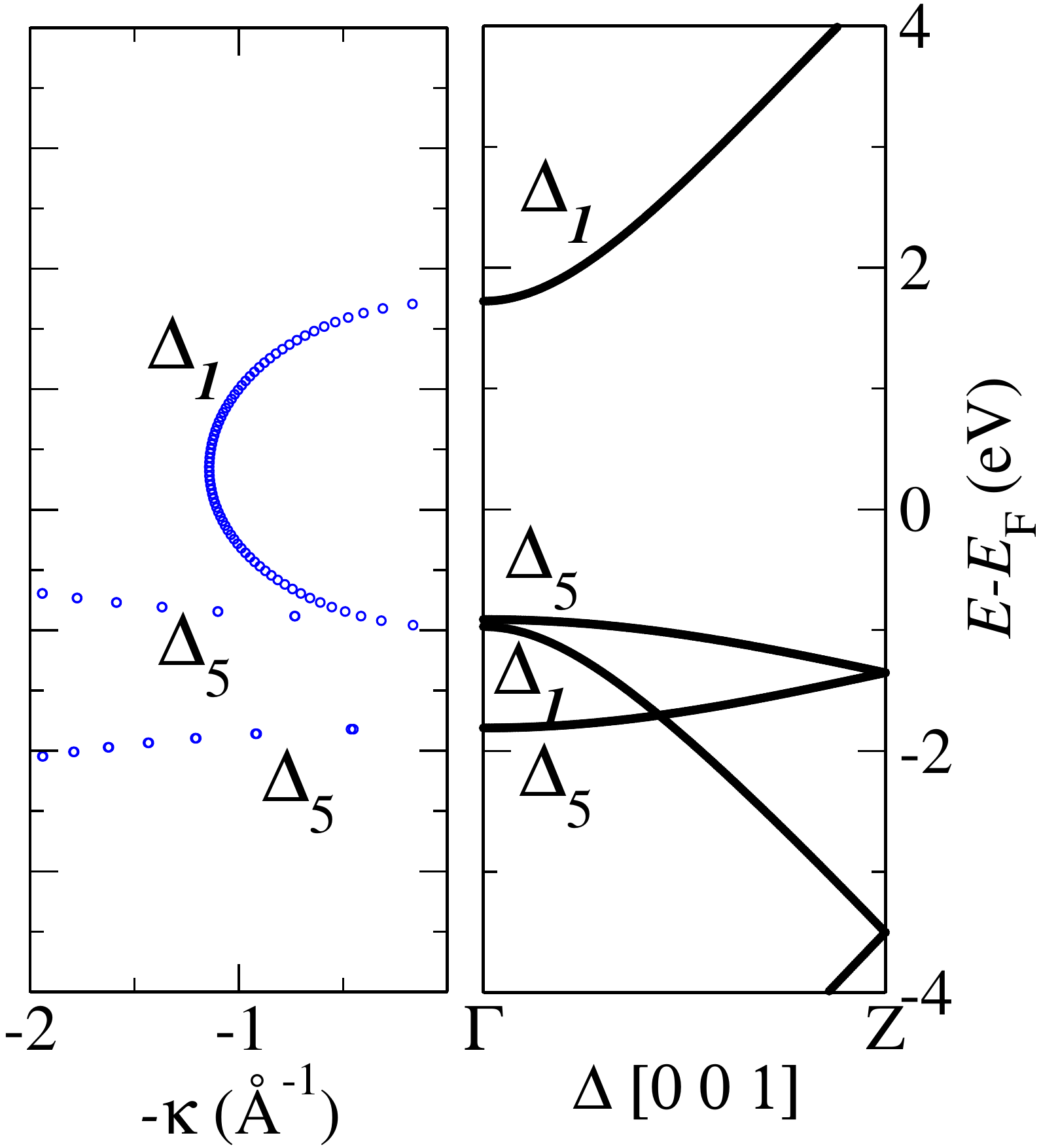}
\caption{Real (right-hand side panel) and complex (left-hand side panel) band structure of ZnO calculated at the 
$\Gamma$ point in the 2D transverse Brillouin zone. The symmetry labels, $\Delta_n$  where n  $\in$[1,5], have
been described in the text and the energy is measured from the Fermi energy $E_\mathrm{F}$.} 
\label{Fig4}
\end{figure}

The real and complex band structures at $\mathbf{k}_\parallel=\Gamma$, for both AlN and ZnO are presented in Fig.~\ref{Fig3} 
and Fig.~\ref{Fig4}, respectively. In both cases there is a continuous semi-circular band that connects the conduction band
bottom to the valence band top across the gap. No low-lying spurious flat bands are observed in our calculations, at variance
to what may happen with non-orthogonal basis sets~\cite{Bosoni2022,DiCarlo}. For both insulators such semi-circular band
is characterised by $\Delta_1$ symmetry, a feature expected since the conduction band bottom is mainly $s$-like. 
Notably, there is another band at the valance band maximum with $\Delta_5$ symmetry. However, this has a rather large 
imaginary wave-vector (decay rate) and will contribute little to the transport, unless the Fermi level of the junction is pinned very 
close to the top edge of the valence band. In that case both the $\Delta_1$ and $\Delta_5$ symmetry states will compete 
to the transmission.

\subsection{Symmetry of the magnetic electrodes}
We now perform the same symmetry analysis for the real band structures of the ferromagnetic electrodes. Given the structure
of our proposed MTJs the relevant direction is [111]. Ideally, the best situation we can encounter is that where there is only
one spin sub-band crossing the Fermi level with the symmetry matching that of the most transmissive complex band, $\Delta_1$
in this case. In such case only one spin channel (either up or down) can be transmitted with high probability, so that the junction 
effectively behaves as a half metal with an almost 100\% spin-polarised current in the parallel configuration, and a magnetoresistance 
ratio increasing exponentially with the barrier thickness. This favourable band alignment is encountered for a band with $\Delta_1$
symmetry in Fe/MgO~\cite{Butler,Mathon} and Fe/HfO$_2$~\cite{shukla} MTJs along the [001] transport direction. 
\begin{figure}[htp]
  \centering
 \includegraphics[width =5.cm]{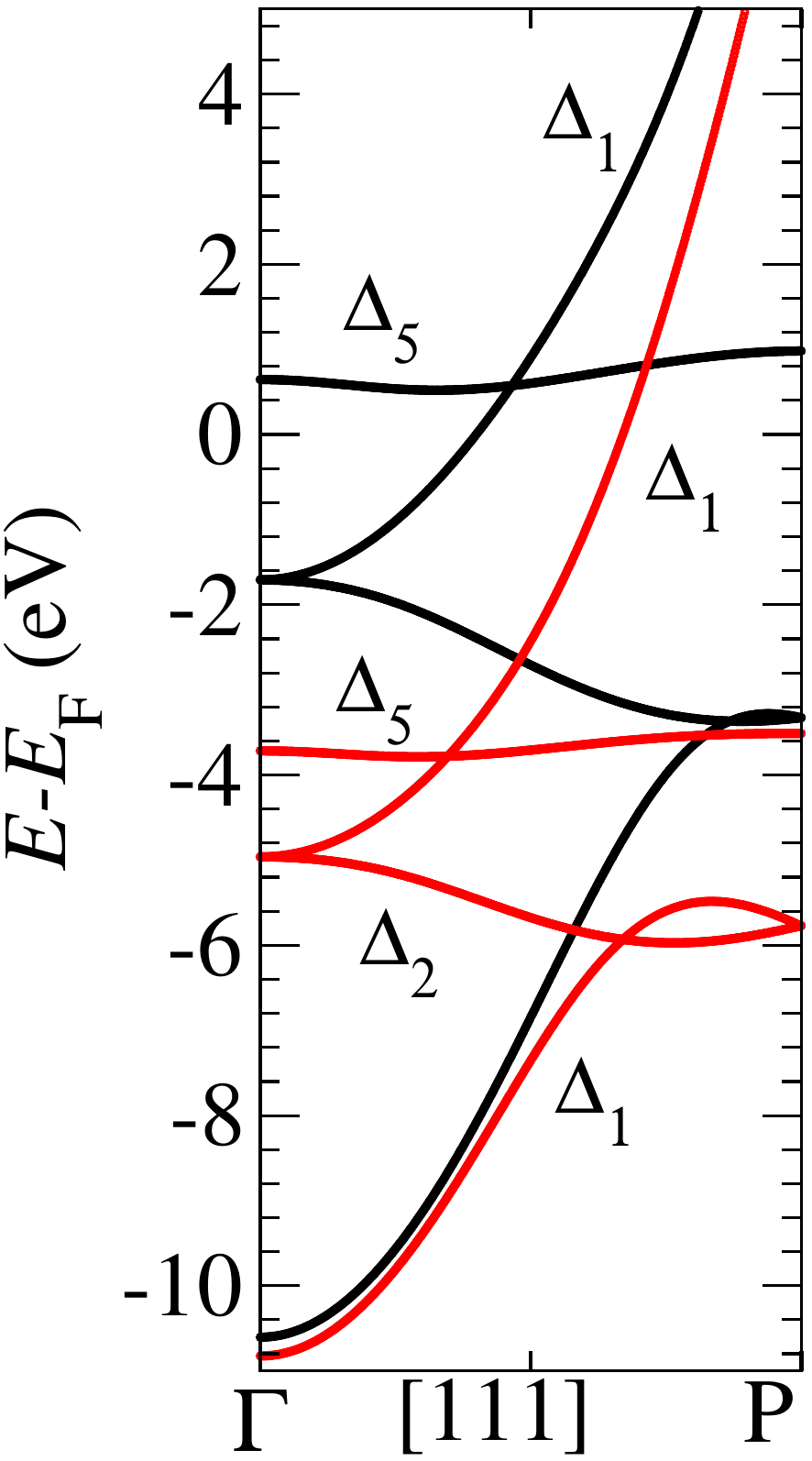}
  \caption{Real band structure of {\it bcc} Co is plotted along the [111] direction (the direction of transport). The majority spin sub-band 
  is in red and the minority one in black.}
  \label{Fig6}
\end{figure}

\begin{figure}[htp]
  \centering
\includegraphics[width =5.cm]{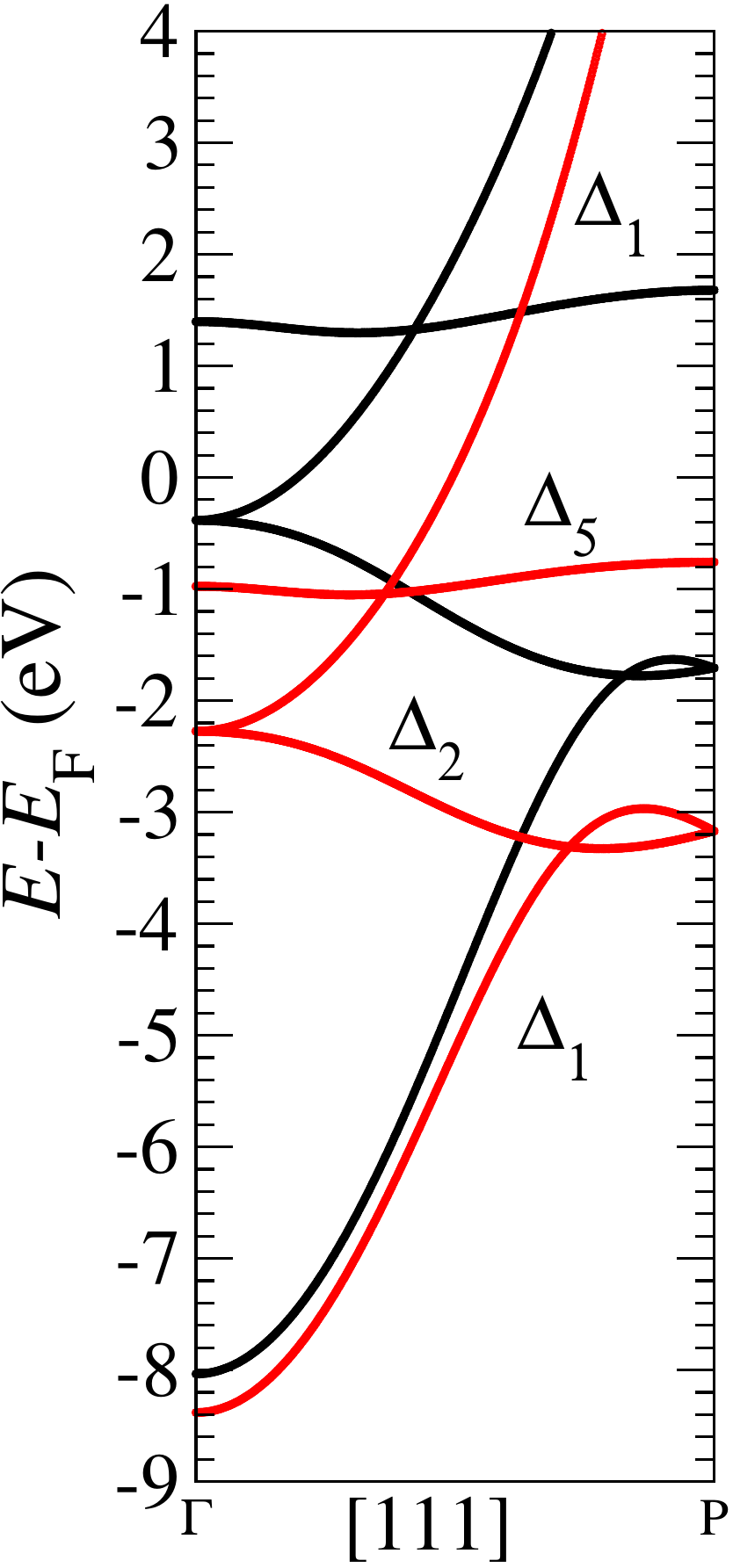}
  \caption{Real band structure of {\it bcc} Fe plotted along the [111] direction (the direction of transport). The majority spin sub-band is 
  in red and the minority one in black.}
  \label{Fig7}
\end{figure}

The real band structure of \textit{bcc}-Co and  \textit{bcc}-Fe are here plotted along the [111] direction in Fig.~\ref{Fig6} and 
Fig.~\ref{Fig7}, respectively. Unfortunately we find that in both ferromagnets, the $\Delta_1$ symmetry is available at Fermi 
energy along [111] for both spins. This means that spin-filtering across the $\Delta_1$ complex band is unlikely, since the 
difference between the two spin sub-bands remains only in the details of the band curvature. Fe seems to offer the most
favourable condition, since the $\Delta_1$ band-edge for the minority sub-band is only about 0.5~eV below the Fermi level. 
This may suggest that under moderate bias conditions there will be regions in the energy window where only one spin
can be transmitted. Together with the very broad $\Delta_1$ bands we also observe two flat bands with $\Delta_5$ and
$\Delta_2$ symmetry, which distribute across $E_\mathrm{F}$ depending on the compound and the spin. The most relevant
for transport appears to be the minority $\Delta_5$ for Co, which is positioned rather close to $E_\mathrm{F}$.

\subsection{Tunnel magnetoresistance}
\begin{figure}[htp]
	\centering
	\includegraphics[width =8.0cm]{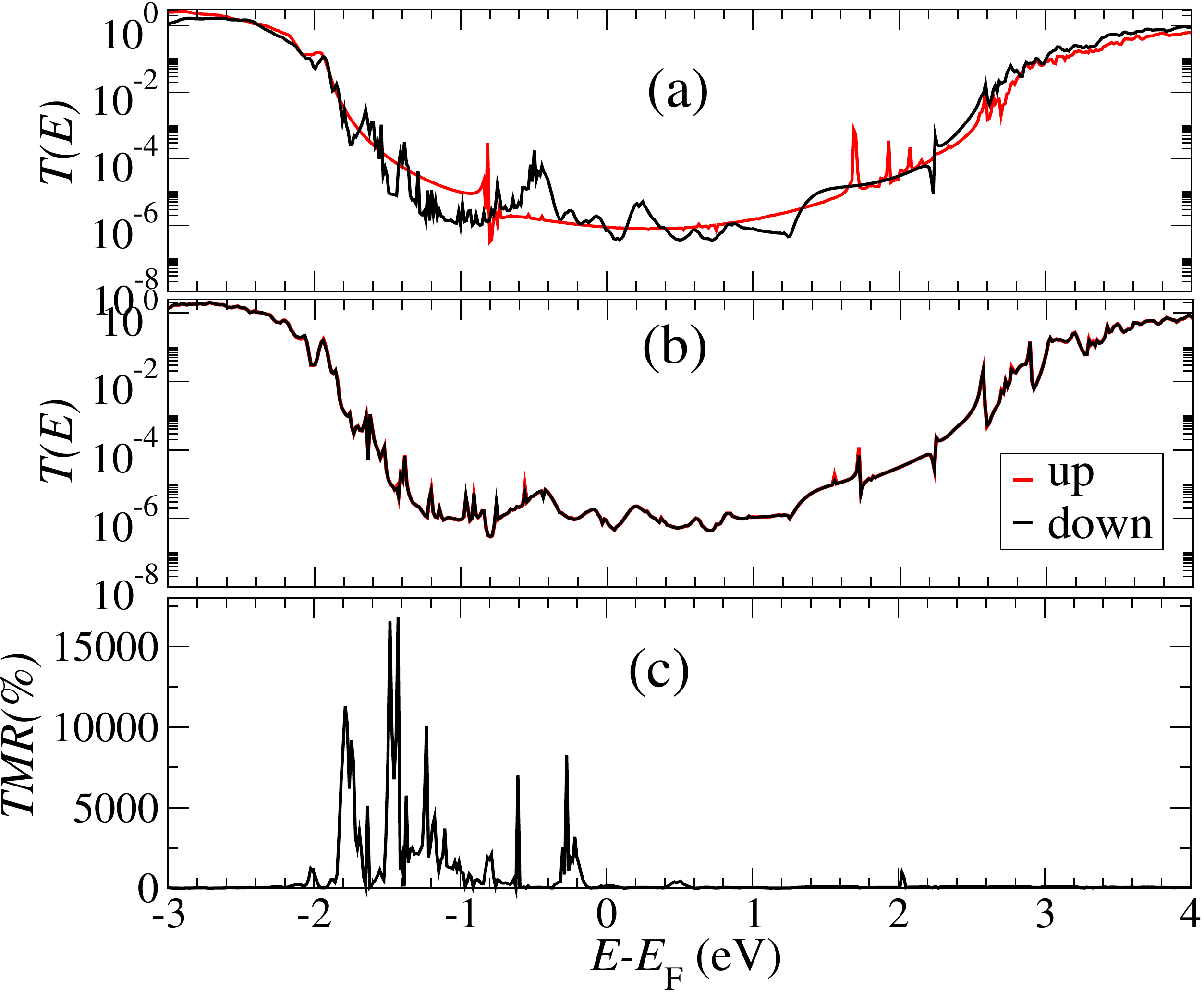}
	\caption{(Colour on line) Transmission coefficient as a function of energy for the Co/AlN/Co MTJ. 
		The parallel and antiparallel configurations are plotted in panel (a) and (b) respectively. $T(E)$ for the majority (minority) 
		spins is plotted in red (black). For the antiparallel case the spin direction is set by the right-hand side electrode. The transmission 
		coefficient is plotted on a logarithmic scale. In the lower panel (c) we present the calculated zero-bias TMR as a function of energy 
		in the same energy window of the transmission coefficients.}
	\label{Fig8}
\end{figure}

Finally, we turn our attention towards the transmission coefficients and the TMR of the proposed MTJs. Let us begin
with AlN-based junctions. Fig.~\ref{Fig8} shows $T(E)$ for both the spin channels ($\uparrow, \downarrow$) in the parallel 
and anti-parallel configuration of the Co/AlN/Co MTJ. As expected, $T(E)$ drops drastically in an energy region approximately 
6~eV wide, which corresponds to the calculated AlN band-gap (note that the transmission coefficient is plotted on a log scale). 
The Fermi level of the junction is positioned at about 2~eV above the AlN valence band so that the MTJ at low bias is deep in 
the tunnelling regime and away from any band edge. This means that there is little contribution to the transmission from
any band with symmetry different from $\Delta_1$. Such observation is corroborated by the shape of $\log[T(E)]$ as a function
of $E$, which resembles closely the complex band of AlN (see Fig.~\ref{Fig3}).

\begin{figure}[htp]
	\centering
	\includegraphics[width =8.0cm]{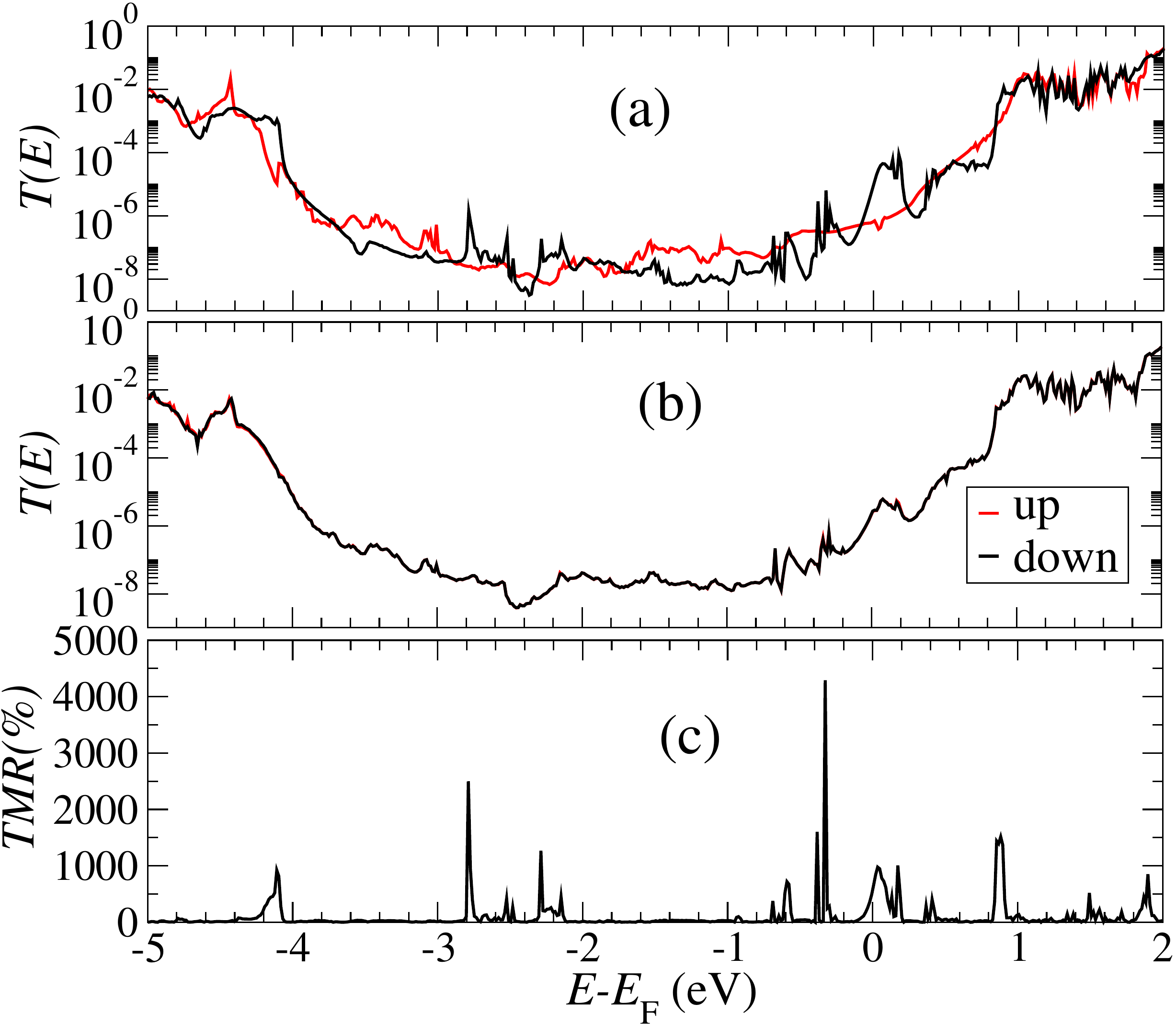}
	\caption{(Colour on line) Transmission coefficient as a function of energy for the Fe/AlN/Fe MTJ. 
		The parallel and antiparallel configurations are plotted in panel (a) and (b) respectively. $T(E)$ for the majority (minority) 
		spins is plotted in red (black). For the antiparallel case the spin direction is set by the right-hand side electrode. The transmission 
		coefficient is plotted on a logarithmic scale. In the lower panel (c) we present the calculated zero-bias TMR as a function of energy 
		in the same energy window of the transmission coefficients.}
	\label{Fig9}
\end{figure}

\begin{figure}[htp]
	\centering
	\includegraphics[width =8.5cm]{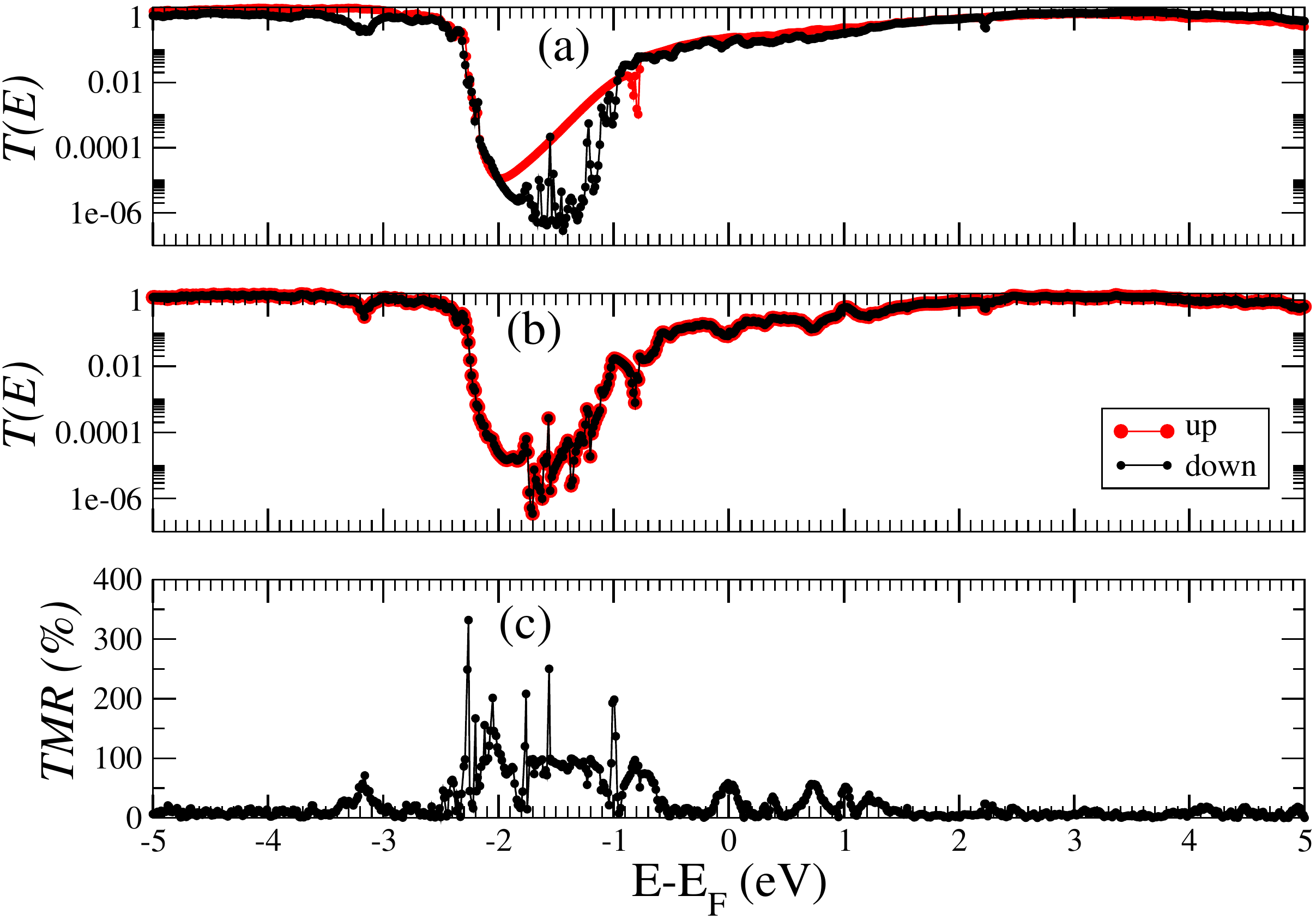}
	\caption{(Colour on line) Transmission coefficient as a function of energy for the Co/ZnO/Co MTJ. 
		The parallel and antiparallel configurations are plotted in panel (a) and (b) respectively. $T(E)$ for the majority (minority) spins 
		is plotted in red(black). For the antiparallel case the spin direction is set by the right-hand side electrode. The transmission coefficient 
		is plotted on a logarithmic scale. In the lower panel (c) we present the calculated zero-bias TMR as a function of energy 
		in the same energy window of the transmission coefficients.}
	\label{Fig10}
\end{figure}

\begin{figure}[htp]
	\centering
	\includegraphics[width =8.0cm]{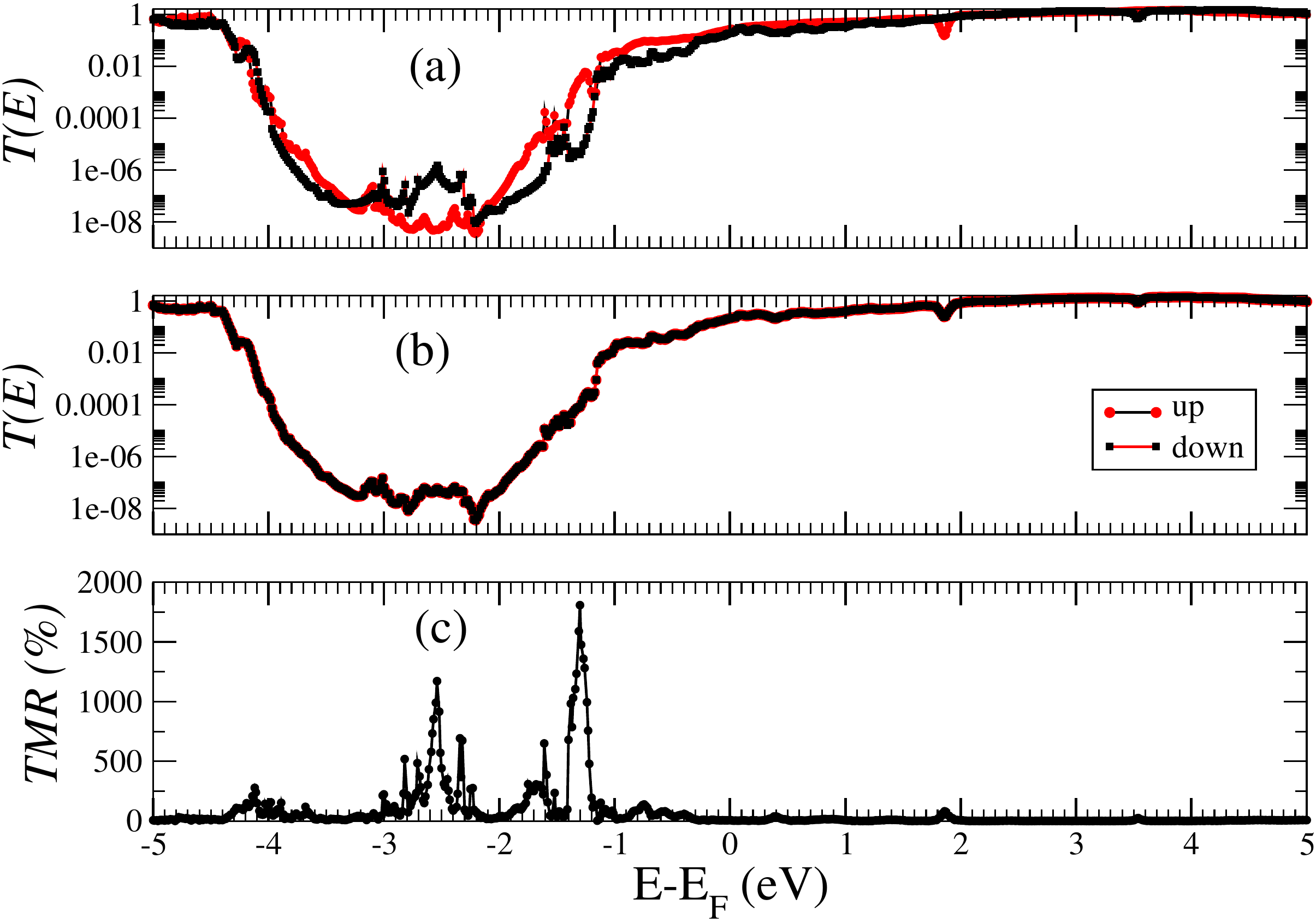}
	\caption{(Colour on line) Transmission coefficient as a function of energy for the Fe/ZnO/Fe MTJ. 
		The parallel and antiparallel configurations are plotted in panel (a) and (b) respectively. $T(E)$ for the majority (minority) spins 
		is plotted in  red (black). For the antiparallel case the spin direction is set by the right-hand side electrode. The transmission coefficient 
		is plotted on a logarithmic scale. In the lower panel (c) we present the calculated zero-bias TMR as a function of energy 
		in the same energy window of the transmission coefficients.}
	\label{Fig11}
\end{figure}

Panel (a) of Fig.~\ref{Fig8} presents the $T(E)$ of Co/AlN/Co in the parallel spin configuration. One can notice that by large the transmission
is similar for the two spins. This is expected from the fact that the highest $\Delta_1$ band-edge for Co along [111] is about 2~eV below
$E_\mathrm{F}$ and corresponds to the minority spin (the one for the majority is well below 4~eV), meaning that across the bandgap
region both $\Delta_1$ spin sub-bands are available to transport. Differences, however, appear as well, with two energy regions where the transmission 
is dominated by one spin only. This happens in the ranges $E_\mathrm{F}-1.5$  to $E_\mathrm{F}-1$ eV for the majority $\Delta_1$ sub-bands,
$E_\mathrm{F}-1$ to $E_\mathrm{F}+0.5$ eV for the minority $\Delta_1$,  $E_\mathrm{F}+0.5$ to $E_\mathrm{F}+1.5$ eV for the majority 
$\Delta_1$, and finally between $E_\mathrm{F}+1.5$ and  $E_\mathrm{F}+2$ eV the minority $\Delta_1$ sub-band dominates. 
As expected, the flat  minority $\Delta_5$  sub-bands with nearly zero band-width has negligible contribution to transmission around 
$E_\mathrm{F}$.

The resulting TMR as a function of energy is then plotted in the lower panel of Fig.~\ref{Fig8} for an energy window of 3 eV around 
$E_\mathrm{F}$. As expected from our transmission coefficient analysis we find a significant TMR  in a region of 2~eV below the Fermi 
level. The maximum value of $\sim$15,000\%  is reached at $E=E_F-1.5$~eV. This  is indeed larger than what is expected from  the simple 
product of the spin-polarized DOS as from Julliere's analysis, indicating  that some spin filtering effect is at work (note that this energy window
is relatively near to the minority $\Delta_1$ band-edge of Co). Unfortunately, this effect takes place far away from the Fermi level so that it will 
be accessible only at extremely large bias voltages.

A similar situation is found for the Fe/AlN/Fe MTJ, whose transmission coefficients and TMR are presented in Fig.~\ref{Fig9}. Also, in this
case panel (a) helps us in understanding the tunnelling process. This time one expects a similar transmission for both spins, down to
about 0.5~eV from the Fermi level, a position corresponding to the minority $\Delta_1$ edge of Fe. In fact, a relatively sharp peak in the 
TMR is found around that energy [see panel (c)], together with some other peaks scattered across the energy window investigated. 
Notably, we did not find a large energy window where the TMR is consistently large, not even for $E<E_\mathrm{F}-0.5$~eV. This seems
to suggest that the $\Delta_1$ transmission away from the $\Gamma$ point contributes sensibly to the tunnelling current by reducing
the spin polarization. Note that for both the AlN-based junctions the transmission in the anti-parallel configuration is spin-independent,
owning to the inversion symmetry of the junction.
  
Finally we move to the ZnO-based junctions, whose transport quantities, $T(E)$ and TMR($E$), are presented in Fig.~\ref{Fig10}
and Fig.~\ref{Fig11}, respectively for Co and Fe electrodes. The main feature of these two junctions is that the Fermi energy
just pins at the conduction band bottom, so that the transport is essentially metallic. In this situation, clearly the spin-filtering 
effect is not at play and the magneto-transport response of the device is determined by the electrodes density of state and
the bonding at the interface. As a result, the TMR at the Fermi level remains always below 100\%, namely it is relatively limited. 
Furthermore, since we are away from the tunnelling limit, we do not expect that the magnetoresistance will depend drastically 
on the barrier thickness. 

Interestingly, for both Co/ZnO/Co and Fe/ZnO/Fe MTJs we find energy regions corresponding to the ZnO bandgap, where
the transmission indeed displays a spin filtering effect. For instance, in Co/ZnO/Co there is a clear dominance of the 
majority $\Delta_1$ transmission in an energy range going from $E_\mathrm{F}-2$~eV to $E_\mathrm{F}-1$~eV,
where the Co minority band instead has a strong $\Delta_2$ character (this is a low-conducting band). As a consequence, 
a substantial TMR is found over this energy range. Similar behaviours are also found for Fe/ZnO/Fe, whose 
transmission spectrum for the parallel configuration [fig.\ref{Fig11}(a)] presents several energy sections with 
a spin sub-band dominating over the other. This is for instance the case in the interval 
$E_\mathrm{F}-3\:\mathrm{eV}<E<E_\mathrm{F}-2\:\mathrm{eV}$, where the high-transmission $\Delta_1$ minority 
band coexists with the majority low-transmission $\Delta_2$. An opposite situation is found for 
$E_\mathrm{F}-2\:\mathrm{eV}<E<E_\mathrm{F}-1\:\mathrm{eV}$, where now the majority $\Delta_1$ 
dominates over the transmission of the minority $\Delta_2$. Unfortunately, these energy regions are not accessible
in practice by the unfavourable pinning of the Fermi energy at the bottom of the ZnO conduction band.

\section{Conclusion}

In summary, we have explored  the possibility of using  display materials AlN and  ZnO as tunnel barriers in novel 
magnetic tunnel junctions. Both of these compounds are currently used in the microelectronic industry, so that 
their MTJs have the potential to be integrated in hybrid memory/logic components or spin-polarized-based display 
devices. When incorporated in an MTJ, we found that both AlN and ZnO change their crystal structure from bulk 
wurzite to a layered planar one. This, however, does not affect drastically their electronic structure and
insulating nature. A complex-band structure analysis has allowed us to identify the dominant symmetry of the 
tunnelling states. In both cases the smallest wave-function decay coefficient is found at the $\Gamma$ point
in the 2D transverse Brillouin zone, although for AlN other high symmetry points present low transmission as well. 
When looking along the transport direction we find that the symmetry of the complex band structure is $\Delta_1$, 
as suggested by the $s$-type character of the insulators' conduction band and by the fact that the lowest complex 
band structure connects directly across the gap. 

We have then investigated four potential MTJs, namely Co/AlN/Co, Fe/AlN/Fe, Co/ZnO/Co and Fe/ZnO/Fe, where
the insulators are oriented along the [0001] direction and the metals along [111]. When ZnO is used as tunnel barrier
the Fermi level pins at the bottom of the conduction band and the transport is therefore metallic. In this case the 
TMR is not determined by spin filtering and it remains limited. This is unfortunate, since deep in the band-gap
spin filtering is active and robust. The situation is more favourable for AlN-based junctions, in particular when
the electrode is Fe. In this case, in fact, there is an energy window around the Fermi level where the majority 
transmission is $\Delta_1$ dominated, but the same band is not available in the minority sub-band. As a consequence
a large TMR at $E_\mathrm{F}$ is found. Our work thus shows that it is possible to potentially achieve large TMR's 
even in junctions with $C_6$ planar symmetry, although new magnetic electrodes may turn out to 
be more suitable than the simple Co and Fe investigated here.


\textit{Acknowledgments}.
This work is supported by the National Research Foundation of Korea (Basic Science Research 
Program: 2021R1A2C1006039), and KISTI (KSC-2021-CRE-0188). SS thanks the Irish Research
Council for financial support (IRCLA/2019/127).


\end{document}